\documentclass{icrc2009}

\usepackage{graphicx}   
\usepackage{caption}    
\usepackage[font=footnotesize]{subfig} 
\usepackage{fixltx2e}
\usepackage{url}

\newcommand{\shorttitle}[1]%
{\markboth{Proceedings of the 31\MakeLowercase{$^{st}$} ICRC, {\L}\'{o}d\'{z} 2009}{#1} }
\newcommand{\etal}{\MakeLowercase{\textit{et al. }}} 

\begin{document}
\title{A synchrotron self-Compton model for the VHE gamma-ray emission from Cen\,A}

\author{\IEEEauthorblockN{J.-P.~Lenain\IEEEauthorrefmark{1},
    M.~C.~Medina\IEEEauthorrefmark{1},
    C.~Boisson\IEEEauthorrefmark{1},
    G.~E.~Romero\IEEEauthorrefmark{2},
    H.~Sol\IEEEauthorrefmark{1} and
    A.~Zech\IEEEauthorrefmark{1}}\\
\IEEEauthorblockA{\IEEEauthorrefmark{1}LUTH, Observatoire de Paris, CNRS, Universit{\'e} Paris Diderot; 5 Place Jules Janssen, 92190 Meudon, France\\
Email: \url{jean-philippe.lenain@obspm.fr}, \url{clementina.medina@obspm.fr}}
\IEEEauthorblockA{\IEEEauthorrefmark{2}Instituto Argentino de Radioastronom\'ia (CCT La Plata-CONICET), C.C.5, (1984) Villa Elisa,\\Buenos Aires, Argentina}}

\shorttitle{J.-P. Lenain~\etal An SSC model for Cen\,A}
\maketitle

\begin{abstract}
    The H.E.S.S. experiment has reported the detection of very high energy (VHE:~$E>100$\,GeV) $\gamma$-ray emission from the well-known radio-galaxy Cen\,A. Following this discovery, we refine the previous multi-blob synchrotron self-Compton (SSC) model applied to the multi-wavelength emission from this source using the new H.E.S.S. data. The prediction of the VHE $\gamma$-ray level of emission for Cen\,A presented in Lenain et~al. paper agrees very well in the view of this recent data. Further VHE observations of Cen\,A might detect variability, which would comfort our inner jet modeling. The next generation of imaging atmospheric \v{C}erenkov telescopes, will help to reveal possible multiple sources of VHE emission in the complex structure of Cen\,A.
  \end{abstract}

\begin{IEEEkeywords}
 active galaxies, galaxies: individual: Cen\,A, non-thermal radiation.
\end{IEEEkeywords}

\section{Introduction}
  The H.E.S.S. experiment has recently reported the detection of VHE $\gamma$-ray emission from the most nearby radio-galaxy: Cen\,A \cite{2009arXiv0903.1582H}. This elliptical galaxy presents strong activity in a wide spectral range from radio to high energy $\gamma$-rays. It is a typical radio-loud AGN viewed with a jet at large angle from the line of sight.

  The reported detection is of high importance as it establishes radio-galaxies as VHE $\gamma$-ray emitters. Cen\,A is the second non-blazar AGN discovered at VHE, after the HEGRA detection of VHE  $\gamma$-rays from M\,87 between 1998 and 1999 \cite{2003A+A...403L...1A} \cite{2004NewAR..48..407B}, confirmed by  H.E.S.S. observations in 2006 \cite{2006Sci...314.1424A}.

  The best-fit position obtained by H.E.S.S does not support the giant radio lobes as VHE $\gamma$-ray emitting zone. The emission direction seems to be compatible with the radio core and the inner jet region. The differential photon spectrum follows a power law  with a photon index of $2.7 \pm 0.45_{\rm stat} \pm 0.1_{\rm sys}$ and no flux variability was found in the period of observation.

  In a previous publication \cite{2008A+A...478..111L}, a new synchrotron self-Compton (SSC) radiative scenario was developed to explain the multi-wavelength emission of misaligned blazar-like objects, such as M\,87 and FR\,I radio-galaxies. In this framework, three objects were investigated, namely Cen\,A, PKS\,0521$-$36 and 3C\,273, with prediction of their VHE $\gamma$-ray emission.

  In this paper we present a more detailed update on the results of this model for Cen\,A with the addition of the new data from H.E.S.S. and {\it Fermi}/LAT. Section~\ref{sec:CenA} contents a brief compilation of observational data on Cen\,A. A short description of the multi-blob SSC scenario developed in \cite{2008A+A...478..111L} is given in Section~\ref{sec:model}. Following, the implications of the model results on the description of this source are summarized in Section ~\ref{sec:results}. Finally, some discussions about the model parameters, limits and future developments are given in Section~\ref{sec:discussion}.

\section{Cen\,A \label{sec:CenA}}

Cen\,A is the closest FR\,I radio-galaxy ($3.8$\,Mpc, \cite{2004A+A...413..903}) and its proximity makes it uniquely observable among such objects, even though its bolometric luminosity is not large by AGN standards. It is very active at radio wavelength, presenting an extremely rich jet structure. We can distinguish in this structure two components: inner jets at a kpc scale and giant lobes which covers $\sim$8$^\circ$ in the sky. A detailed description of its radio morphology can be found in \cite{1989AJ.....98...27M}. The inner kpc jet has also been detected in X-rays \cite{2002ApJ...569...54K} with a structure of knots summed to diffuse emission. Recently, \cite{2009arXiv0901.1346C} reported the detection of non-thermal X-ray emission from the shock of the southwest inner radio lobe from deep {\it Chandra} observations.

 The supermassive black hole at the center of the active galaxy has an estimated mass of about $5.5 \times 10^7 M_\odot$ \cite{2009MNRAS...394..660C} to $10^8 M_\odot$ \cite{2005AJ....130..406S}. The stellar body associated with the radio source is an elliptical galaxy (NGC\,5128) with a twisted disk of dust, gas and young stars which obscures the central engine at optical wavelengths. This dust lane and the extended shell structures detected in long exposure optical images are thought to be the remnant of a recent merging process between a massive elliptical galaxy and a spiral one \cite{1992MNRAS.257..689T}.

 The inner part of Cen\,A shows a circumnuclear disk of about 400 pc diameter and a central cavity of about 90 pc. This disk is emitting in the (sub) millimeter range. Against this emission, a variety of molecular absorption lines are seen \cite{1990A+A...227..342I}. In addition, some evidence for a 40 parsec diameter disk (2.5 arcsec) of thin ionized gas centered on the nucleus of Centaurus A has been found in the light of Pa$\alpha$ using the NICMOS instrument on board {\it HST} \cite{1998ApJ...499L.143S}. This disk is not perpendicular to the jet and thus it is not directly associated with the accretion disk of the black hole.

 After the early detection in X-ray was made during a rocket flight \cite{1972ApJ...171L..45L}, between 1991 and 1995, Cen\,A was observed by the Compton Gamma Ray Observatory ({\it CGRO}) and all its instruments from MeV to GeV energies (\cite{1992AIPC..254..348G,1993AIPC..280..473P,1998A+A...330...97S,1995ApJS..101..259T}). These observations revealed a peak in the spectral energy distribution (SED) in $\nu F_\nu$ representation at $\sim0.1$\,MeV with a maximum flux of about  $10^9$\,erg\,cm$^{-2}$\,s$^{-1}$ \cite{1998A+A...330...97S}. At this time it has exhibited large X-ray variability \cite{1996A+A...307..708B}. Some soft gamma-ray variability was also reported (\cite{1998A+A...330...97S,1996A+A...307..708B}) but \cite{1999APh....11..221S} found that EGRET flux was stable during the whole period of {\it CGRO} observations.

 In 1999 the new {\it Chandra} X-ray Observatory took images of Cen\,A with unprecedented resolution. More than 200 point X-ray sources with an integrated luminosity of $L_X > 10^{38}$\,erg\,s$^{-1}$ can be identified in those images \cite{2001ApJ...560..675K}.

 Before the discovery reported by H.E.S.S., a tentative detection of Cen\,A  at VHE during a giant X-ray outburst in the 1970's was reported by \cite{1975ApJ...197L...9G}. Subsequent VHE observations made with the Mark III \cite{1990A+A...228..327C}, JANZOS \cite{1993APh.....1..269J}, CANGAROO \cite{2007ApJ...668..968K}, and H.E.S.S. \cite{2005A+A...441..465A} experiments resulted in upper limits.

 Cen\,A was early proposed by \cite{1996APh.....5..279R} as a possible source of UHE cosmic rays. Recently, the Pierre Auger Collaboration reported the existence of anisotropy on the arrival directions of UHE cosmic rays \cite{2007Sci...318..938T}, remarking that at least 2 of this events can be correlated with the Cen\,A position ($3^{\circ}$ circle). Further works have claimed that there are several events that can be associated with Cen\,A and its big radio lobes (\cite{2008JETPL..87..461G,2008PhyS...78d5901F}) but this correlation is until now statistically weak.

 Finally, the last news about Cen\,A are coming from {\it Fermi}/LAT which has detected it in the three first months of survey with a signal significance above 10$\sigma$ \cite{2009arXiv0902.1559A}.

\section{The multi-blob SSC model for misaligned blazar-like objects \label{sec:model}}

We summarize here the principles of the multi-blob SSC model developed by \cite{2008A+A...478..111L}. The VHE $\gamma$-ray emitting region is assumed to consist in several small blobs traveling through the extended jet. The radiation occurs at $\sim 100 r_g$ (where $r_g$ is the gravitational radius of the central black hole) from the central engine, just beyond the Alfv\'en surface as found in magnetohydrodynamic models \cite{2006MNRAS.368.1561M}, in the broadened jet formation region in the innermost part of the jet. The cap of small blobs in the inner jet introduced in \cite{2008A+A...478..111L} should be seen as a sketch to describe an inhomogeneous turbulent flow crossing a stationary shock. In this case, even for large viewing angles between the jet axis and the line of sight, as it is the case in Cen\,A ($15^\circ$-$80^\circ$) (\cite{2003ApJ...593..169H,1998AJ....115..960T}), some material could move closely aligned to the line of sight, and thus benefit from strong Doppler boosted emission.

In another scenario presented in \cite{2005A+A...432..401G}, the authors develop a model for the VHE $\gamma$-ray emission of BL Lac objects and radio-galaxies based on an structured jet composed by a fast spine surrounded by a slower layer. Depending on the viewing angle between the jet and the line of sight, these components contribute differently to the spectral energy distribution. In the case of M\,87 \cite{2008MNRAS.385L..98T}, this model can also reproduce the data correctly.

\section{Results \label{sec:results}}

We present here an update of the prediction for the VHE flux expected with H.E.S.S. made by \cite{2008A+A...478..111L} (see their Fig.~7). The data sample chosen here is almost the same as in the earlier publication, but it now includes VHE positive detection and {\it Fermi} data. The strongly constraining {\it CGRO}/COMPTEL $\gamma$-ray data are taken from \cite{1998A+A...330...97S}. The {\it RXTE} and {\it INTEGRAL} data were provided by \cite{2006ApJ...641..801R}. From \cite{2000ApJ...528..276M} we obtained the {\it HST}/NICMOS and WFPC2 data, which was carefully dereddened. SCUBA data at 800 $\mu$m was taken from \cite{1993MNRAS...260..844H} while ISO and SCUBA (450 $\mu$m and 850 $\mu$m) data are from \cite{1999A+A...341..667}. The VLA data were acquired from \cite{1983ApJ...273..128}. Evans et al. in \cite{2004ApJ...612..786} report X-ray observations by {\it XMM}-Newton on February 2 and 6, 2001 and by Chandra on May 9 and 21, 2001 with a photon index $\Gamma$ = 2 for the parsec-scale jet component. Data from the Nasa Extragalactic Database are also shown as non-constraining points (\textit{in gray}) for comparison.

\begin{figure*}[th]
    \centering%
    \includegraphics[angle=-90,width=5in]{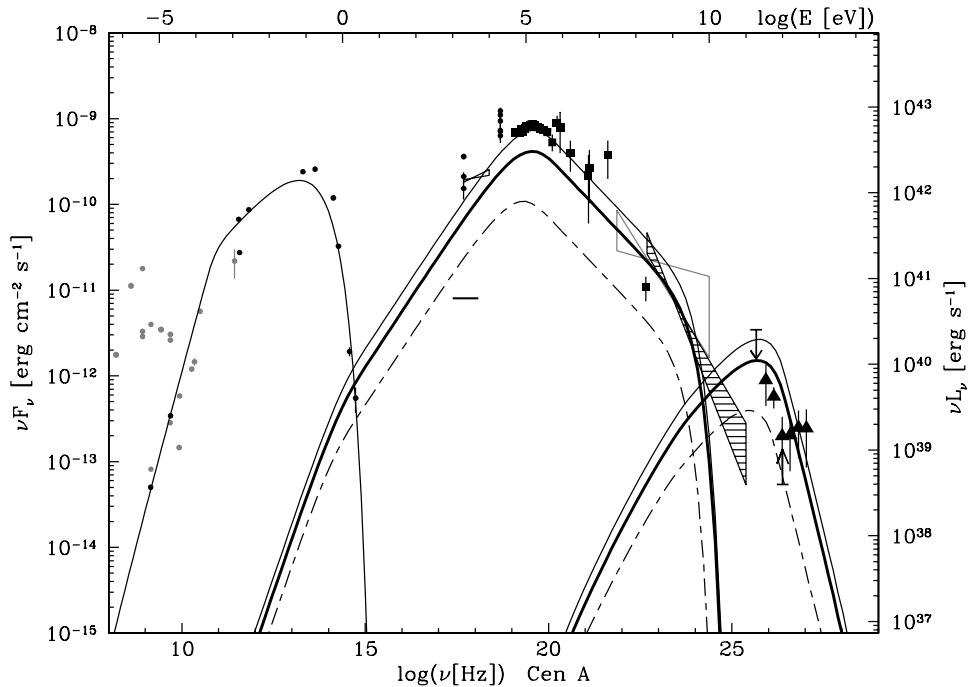}%
      \caption{SED of Cen\,A, along with the H.E.S.S. spectrum in the VHE range (filled triangles). There are also shown the data from {\it CGRO}/COMPTEL \cite{1998A+A...330...97S}, {\it RXTE} and {\it INTEGRAL} \cite{2006ApJ...641..801R}, {\it HST}/NICMOS and WFPC2 \cite{2000ApJ...528..276M}, SCUBA at 800 $\mu$m  from \cite{1993MNRAS...260..844H}, ISO and SCUBA (450 $\mu$m and 850 $\mu$m)\cite{1999A+A...341..667}, the {\it XMM}-Newton and {\it Chandra} spectrum, the {\it Fermi}/LAT spectrum (shaded bow-tie) considering a simple power law and finally, data from the NASA Extragalactic Database (NED) (grey filled circles). The thick black line represents the average result of the multi-blob SSC model, obtained with the parameters given in Table~\ref{tab-Param}. The thin and solid line corresponds to the ``on-blob'' case and the dashed line to the ``inter-blob'' case described in \cite{2008A+A...478..111L}. In the radio and optical ranges, fluxes of the central region of Cen\,A can be reproduced on the average by an extended synchrotron emitting jet.}
\label{fig-SED_CenA}
    \end{figure*}

The {\it Fermi}/LAT collaboration recently published a source list of the bright AGNs \cite{2009arXiv0902.1559A} giving a simple power-law reconstruction of the spectrum. Even though the high energy spectrum of Cen\,A might be more complex, we consider this simple power-law assumption as a first clue of its real spectrum in {\it Fermi} energy range.

  Furthermore, the {\it Fermi}/LAT analysis revealed that the high energy spectrum seems constant over the first three months of scientific operation of the instrument. This constant behavior is also supported by the fact that {\it CGRO} also reported a constant flux from Cen\,A. We thus model together both {\it CGRO} and {\it Fermi}/LAT data.

  Figure~\ref{fig-SED_CenA} gives an update of our multi-blob SSC model to the SED of Cen\,A, slightly modifying the parameters as compared to \cite{2008A+A...478..111L} to account for the recent {\it Fermi}/LAT and H.E.S.S. spectra. We note that a discrepancy appears between the flux normalizations of {\it Fermi}/LAT and H.E.S.S. This difference might be resolved when the spectral points from {\it Fermi}/LAT will become available, however it might also be due to variability in the flux, since the data were not taken simultaneously. The thick black line represents the average result of the SSC, with the parameters given in Table~\ref{tab-Param}. This average is between two extreme scenarios: ``on-blob'' (thin and solid line) and ``inter-blob''(dashed line) (see \cite{2008A+A...478..111L} for more details).

  Considering these recent high energy and VHE results, we favor the second hypothesis presented in \cite{2008A+A...478..111L}, that is the soft $\gamma$-ray data observed by {\it CGRO} and now {\it Fermi} are more likely of synchrotron nature rather than inverse Compton. In Table~\ref{tab-Param} we give both the parameters for this second hypothesis from \cite{2008A+A...478..111L} and from the updated model presented here.

  \begin{table*}[th]
    \caption{Parameters used in the multi-blob SSC model for Cen\,A.}             
    \label{tab-Param}      
    \centering
    \begin{tabular}{c c c}
      \hline\hline
      Parameter            & \cite{2008A+A...478..111L} & this work\\
      \hline
      $\Gamma_b$           & 20 & 15\\
      $\theta$             & $25^\circ$ & $25^\circ$  \\
      $R_\mathrm{cap}$      & 100\,$r_g$ & 160\,$r_g$\\
      $B$                  & 10\,G & 10\,G\\
      $r_b$                & $8.0 \times 10^{13}$\,cm & $9.5 \times 10^{13}$\,cm\\
      $K_1$                & $4.0 \times 10^4$\,cm$^{-3}$ & $3.0 \times 10^3$\,cm$^{-3}$\\
      $n_1$                & 2.0 & 1.7\\
      $n_2$                & 3.5 & 3.9\\
      $\gamma_\mathrm{min}$ & $10^3$ & $10^3$\\
      $\gamma_b$           & $3.5 \times 10^5$ & $4.0 \times 10^5$\\
      $\gamma_c$           & $6.0 \times 10^6$ & $5.0 \times 10^7$\\
      \hline
    \end{tabular}
  \end{table*}

\section{Discussion and perspectives \label{sec:discussion}}

The multi-blob SSC model describes well the observed spectral energy distribution with only a small set of free parameters. The only parameter that differs notably from the standard SSC results for blazars is the relatively high magnetic field of 10~G, which is needed to arrive at the observed flux level. Other solutions are possible due to the inherent degeneracy between the parameters. Even though the muti-blob SSC model achieves a good description for the two misaligned blazars with VHE $\gamma$-rays emission, there is also another leptonic description based on the particle acceleration along rotating magnetic field lines in the magnetosphere of the supermassive black hole, which has been applied successfully to M\,87 and could possibly also work for Cen\,A \cite{2008A+A...479..5}.

A more complete model would need to take into account the existence of hadrons in the jet and their contribution to the observed emission. Current hadronic models investigate the case where emission from hadronic processes dominates the spectral energy distribution at high energies. One possibility is given by the synchrotron proton blazar model \cite{2004A+A...419..89}, in which the VHE emission is dominated by synchrotron radiation from protons in magnetic fields of several tens of Gauss in cases such as M\,87.

In a different scenario, far larger values of the magnetic field are estimated when assuming the emission region to be close to the jet apex \cite{Orellana-Romero}. In such a highly magnetized context, in the innermost part of the jet, synchrotron losses become catastrophic for electrons, and IC interactions lose importance. In this scenario, the resulting photon field becomes a suitable target for photo-pion production by relativistic protons accelerated by shocks in the innermost part of the jet as in models for galactic jets \cite{2008A+A...485..623R}, and protons can achieve energies up to $\sim$10$^{18}$\,eV.


With its intended high resolution and sensitivity, the future Cherenkov Telescope Array (CTA) Observatory will be an ideal instrument to ponder the role played by relativistic protons in the inner part of Cen\,A.

\section*{Acknowledgment}
\addcontentsline{toc}{section}{Acknowledgment}

We would like to acknowledge the support from the GDR-PCHE and LEA-ELGA.

\end{document}